\documentclass[12pt]{iopart}

\usepackage{iopams}
\usepackage{setstack}
\usepackage{graphicx}
\usepackage{epsfig}
\usepackage{epstopdf}
\begin{document}

\title{Microwave Quantum Optics with an Artificial Atom}

\author{Io-Chun Hoi$^{1}$, C.M. Wilson$^{1}$, G\"oran Johansson$^{1}$, Joel Lindkvist$^{1}$, Borja Peropadre$^{2}$, Tauno Palomaki$^{1}$, \& Per Delsing$^{1}$}

\address{$^1$Department of Microtechnology and Nanoscience (MC2), Chalmers University of Technology, SE-412 96 G\"oteborg, Sweden}
\address{$^2$Instituto de F\'isica Fundamental, CSIC, Calle Serrano 113-bis, Madrid E-28006, Spain}
\ead{per.delsing@chalmers.se, chris.wilson@chalmers.se}

\begin{abstract}

We address the recent advances on microwave quantum optics with artificial atoms. This field relies on the fact that the coupling between a superconducting artificial atom and propagating microwave photons in a 1D open transmission line can be made strong enough to observe quantum coherent effects, without using any cavity to confine the
microwave photons. We investigate the scattering properties in such a system with resonant coherent microwaves. We observe the strong nonlinearity of the artificial atom and under strong driving we observe the Mollow triplet. By applying two resonant tones, we also observe the Autler-Townes splitting. By exploiting these effects, we demonstrate two quantum devices at the single-photon level in the microwave regime: the single-photon router and the photon-number filter. These devices provide essential steps towards the realization of an on-chip quantum network.

\end{abstract}

\maketitle

\section{Introduction}



During the last decade, circuit QED based on superconducting circuits has become a promising platform to investigate strong coupling between light and matter as well as enable quantum information processing technology \cite{Schoelkopf,Clarke,Wendin}. Some of the exciting results include the following: strong coupling between a superconducting qubit and a single photon \cite{Wallraff}, resolving photon-number states \cite{Schuster}, synthesizing arbitrary quantum states \cite{Hofheinz}, three-qubit quantum error correction \cite{Reed}, implementation of a Toffoli gate \cite{Fedorov}, quantum feedback control \cite{Vijay} and architectures for a superconducting quantum computer \cite{Mariantoni}. The nonlinear properties of Josephson junctions have also been used to study the  dynamical Casimir effect \cite{Chris2} and build quantum limited amplifiers \cite{Beltran,Bergeal}. 

More recently, theoretical and experimental work has begun to investigate the strong interaction between light and a single atom even without a cavity \cite{Tey,Hwang,Wrigge,Gerhardt}. In this system, the destructive interference between the excited dipole radiation and the incident field gives rise to the extinction of the forward propagating wave for a weak incident field. This experiment was first demonstrated for a single atom/molecule in 3D space, where the extinction of the forward incident wave did not exceed 12$\%$ \cite{Tey,Wrigge}. This is due to the spatial mode mismatch between the incident and scattered waves.

Taking advantages of the huge dipole moment of the artificial atom and the confinement of the propagating fields in a 1D open transmission line \cite{Astafiev1,Hoi,Hoi2,Hoi3,Abdumalikov,Astafiev2,Shen2,Chang}, strong coupling between an artificial atom and propagating field can be achieved. Extinctions in excess of 99\% has been observed \cite{Hoi,Hoi2}. This system represents a potential key component in the field of microwave quantum optics, which is the central scope of this article.

This paper is organized as follows. The elastic scattering properties of the single artificial atom is presented in section 2. Well-known quantum optics effects, such as the Mollow Triplet and Aulter-Townes splitting are presented in section 3. In section 4, we demonstrate two quantum devices based on these effects which operate at the single-photon level in the microwave regime, namely: the single-photon router and the photon-number filter. In section 5, we discuss the possibilities of a quantum network using these devices.

 \begin{figure}
 \includegraphics[width=\columnwidth]{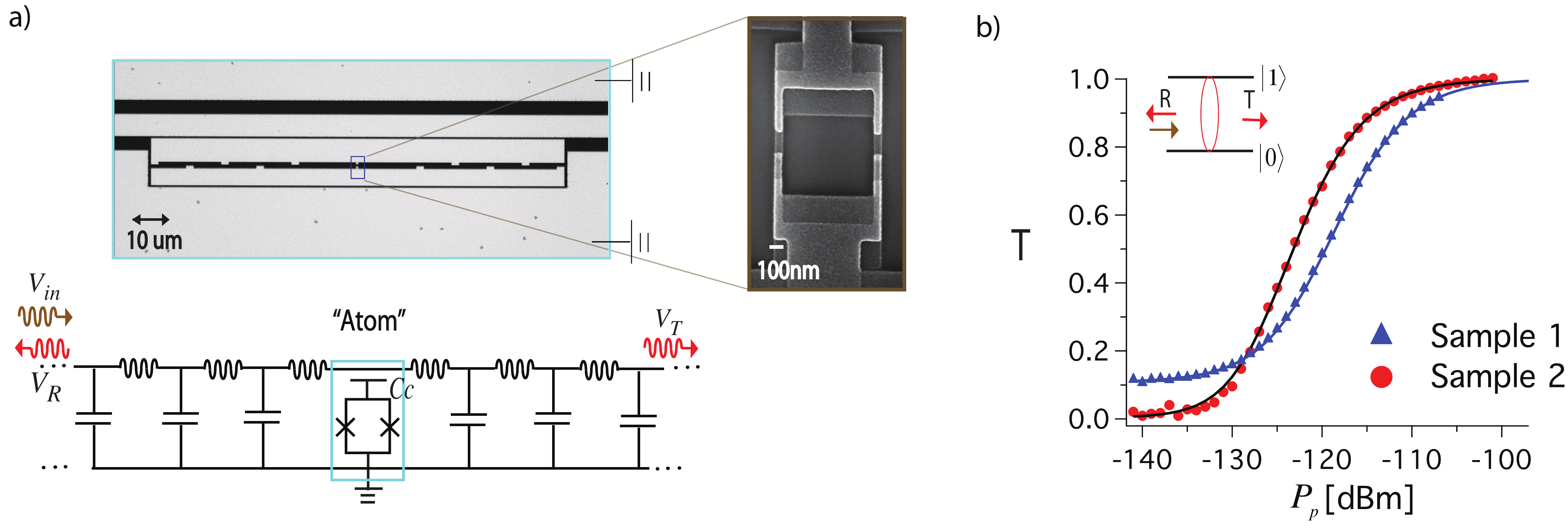}
\caption{(a) Top: A micrograph of the artificial atom, a superconducting transmon qubit embedded in a 1D open transmission line. (Zoom In) Scanning-electron micrograph of the SQUID loop of the transmon. Bottom: the corresponding circuit model. (b) Measured transmittance, $T=|t|^2$, on resonance as a function of the incoming probe power, $P_p$, for sample 1 and 2. At low power, very little is transmitted whereas at high power practically everything is transmitted. (Inset) A weak, resonant coherent field is reflected by the atom.}
\end{figure}

\section{Elastic Scattering}
In Fig. 1a,  a transmon qubit \cite{Koch} is embedded in a 1D open transmission line with a characteristic impedance  $Z_0\simeq 50$ $\Omega$. The 0-1 transition energy of the transmon, $ \hbar\omega_{10}(\Phi)\approx \sqrt{8E_J(\Phi)E_C}-E_C$, is determined by two energies, where $E_C=e^2/2C_\Sigma$ is the charging energy, $C_\Sigma$ is the total capacitance of the transmon,  $E_J(\Phi)=E_{J}|\cos(\pi\Phi/\Phi_0)|$ is the Josephson energy which can be tuned by the external flux $\Phi$, $E_{J}$ is the maximum Josephson energy and $\Phi_0=h/2e$ is the magnetic flux quantum.

With a coherent state input, we investigate the transmission and reflection properties of the field. The input field, transmitted field and the reflected field are denoted as $V_{in}$, $V_{T}$ and $V_{R}$, respectively, indicated in the bottom panel of Fig. 1a.
By definition, the transmission coefficient $t=V_{T}/V_{in} =1+r$. The reflection coefficient, $r$, can be expressed as \cite{Astafiev1},

\begin{eqnarray}
r=\frac{V_{R}}{V_{in}} =-r_0\frac{1-i\delta\omega_p/\gamma_{10}}{1+(\delta\omega_p/\gamma_{10})^2+\Omega_p^2/\Gamma_{10}\gamma_{10}},
\end{eqnarray}
where the maximum reflection amplitude is given by $r_0=\Gamma_{10}/2\gamma_{10}$. $\Gamma_{10}$,$\Gamma_{\phi}$ are the relaxation rate and pure dephasing rate of the 0-1 transition of the atom, respectively. $\gamma_{10}=\Gamma_{10}/2+\Gamma_{\phi}$ is the 0-1 decoherence rate and $\delta\omega_p=\omega_p-\omega_{10}$ is the detuning between the applied probe frequency, $\omega_p$, and 0-1 transition frequency, $\omega_{10}$. We see that both $r_0$ and $\gamma_{10}$ are uniquely dependent on $\Gamma_\phi$ and $\Gamma_{10}$. $\Omega_p$ is the Rabi oscillation frequency induced by the probe, which is proportional to $V_{in}$ \cite{Koch},
\begin{eqnarray}
\Omega_p=\frac{2e}{\hbar}\frac{C_c}{C_{\Sigma}}\left(\frac{E_J}{8E_C}\right)^{1/4}\sqrt{P_pZ_0},
\end{eqnarray}
where $P_p=|V_{in}|^2/2Z_0$ is the probe power. The relaxation process is dominated by coupling to the 1D transmission line through the coupling capacitance $C_c$ (see bottom panel of Fig. 1a) and assuming that photon emission to the transmission line dominates the relaxation, we find $\Gamma_{10}\simeq\omega_{10}^2C_c^2Z_0/(2 C_\Sigma)$.

 \begin{figure*}
 \includegraphics[width=\columnwidth]{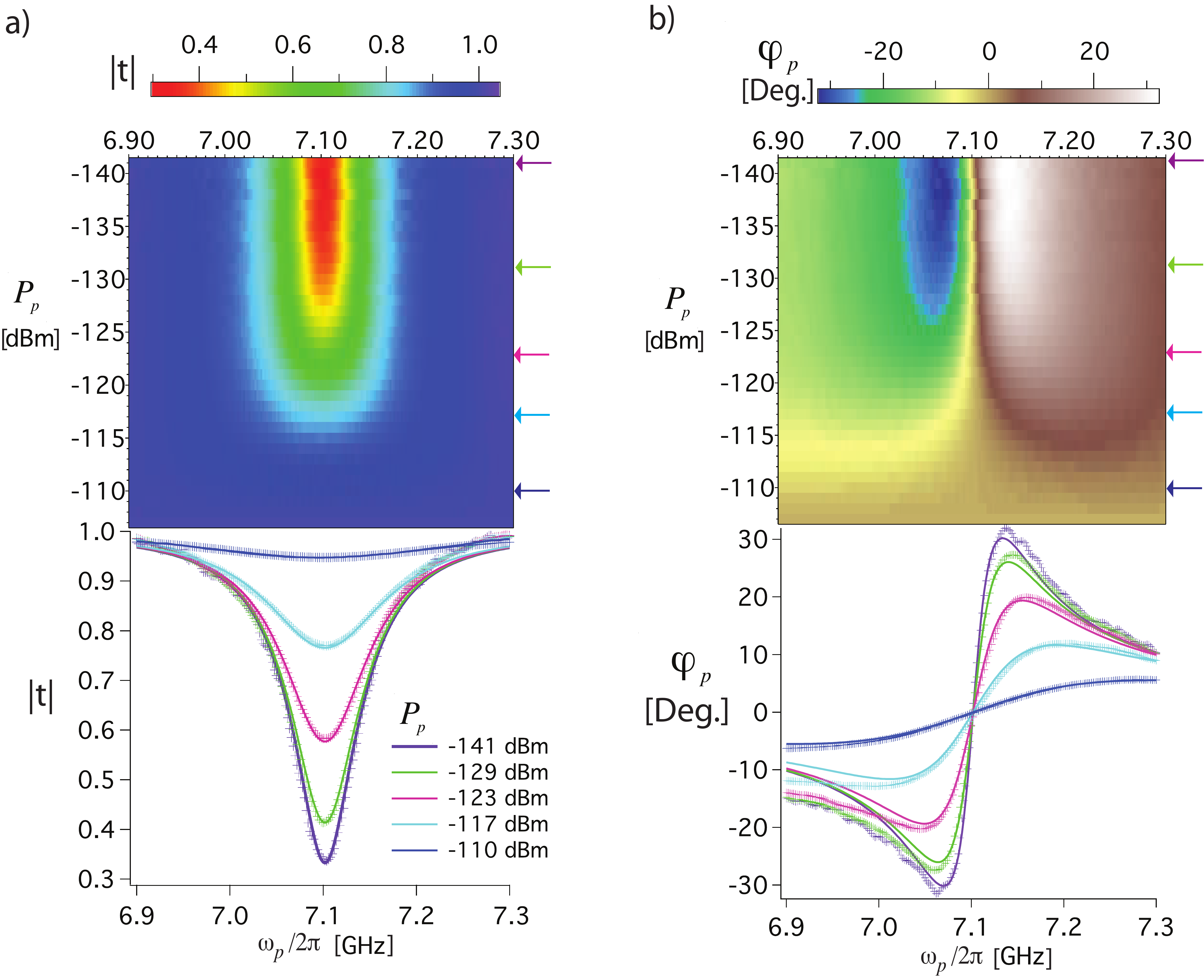}

\caption{t as a function of $P_p$ and $\omega_p$ (sample 1). a) The magnitude response. b) the phase response. Top panel: experimental data. Bottom panel: we show the line cuts for 5 different powers, as indicated by the arrows on the top panel. The experimental data (markers) are fit simultaneously using Eq. 1 (curves). The magnitude response demonstrates the strong coupling between the atom and resonant propagating microwaves, while the phase response shows anomalous dispersion \cite{Astafiev1}.}
\end{figure*}

According to Eq.(1), for a weak $(\Omega_p\ll\gamma_{10})$, resonant probe $(\delta\omega_p=0)$, in the absence of both pure dephasing $(\Gamma_{\phi}=0)$ and unknown relaxation channels, we should see full reflection ($|r|=1$) of the incoming probe field \cite{Zumofen,Chang,Shen2}. In that case, we also have full extinction, $|t|=0$, of the propagating wave. This full extinction (perfect reflection) can be described as a coherent interference of the incoming wave and the scattered wave from the atom. This is what we observe in Fig. 1b, we measure the transmittance, $T=|t|^2$, on resonance as a function of $P_p$ with two samples. We see an extinction in the resonant microwaves of up to $ 90 \%(99 \%)$ for sample 1(2) at low incident probe power, where $\Omega_p\ll\gamma_{10}$. For increasing $P_p$, we see the strong nonlinearity of the atom which becomes saturated by the incident microwave photons. Since the atom can only scatter one photon at a time, at high incident power, $\Omega_p\gg\gamma_{10}$, most of the photons pass the atom without interaction and are thus transmitted. Therefore $|t|$ tends towards unity for increasing $P_p$, consistent with Eq.(1). We define the average probe photon number coming to the transmon per interaction time as, $\left \langle N_p \right \rangle=P_p/(\hbar\omega_p(\Gamma_{10}/2\pi))$.

We measure $t$ as a function of $P_p$ and $\omega_p$. In Fig. 2, the experimental magnitude, $|t|$, and phase response, $\varphi_p$, for sample 1 are shown in a, b, respectively. The top and bottom panels display 2D plots and the corresponding line cuts indicated by the arrows, respectively. For $\left \langle N_p \right \rangle\ll1$, the magnitude response shows the strong extinction of resonant microwaves, up to 70$\%$ in amplitude or $\sim 90\%$ in power (Fig. 1b). The solid curves of Fig. 2 show fits to all magnitude and phase curves simultaneously, with three fitting parameters, $\Gamma_{10}/2\pi=73$\,MHz, $\Gamma_{\phi}/2\pi=18$\,MHz  and $\omega_{10}/2\pi=7.1$\,GHz.  This corresponds to  $C_c=25$\,fF, $\gamma_{10}/2\pi=55$\,MHz and $r_0=0.67$. We find very good agreement between theory and experiment. We also see that $r$ varies as a function of $P_p$ and $\omega_p$, as expected (data not shown).

To further characterize the sample 1, the resonance dip in transmission in Fig. 2a is mapped as a function of $\Phi$ at a weak probe, $\Omega_p\ll\gamma_{10}$ (see Fig. 3a). If we increase $P_p$ to a level such that the 0-1 transition is saturated, two-photon (0-2) transitions occur, as indicated in the grey curve of Fig. 3b. The transition frequency corresponds to $(\omega_{10}+\omega_{21})/2$, where $\omega_{21}$ is the 1-2 transition energy. We use a Cooper Pair Box \cite{Koch} Hamiltonian with 50 charge states to fit the spectrum of the atom. We extract  $E_{J}=12.7$\,GHz, $E_C=590$\,MHz for sample 1. The extracted parameters are summarized in Table 1. Note that one of the Josephson junction is broken in sample 3, therefore, the transition frequency could not be tuned with $\Phi$.

\begin{table}
\begin{tabular}{ccccccccc}
\hline
Sample & $E_J/h$ & $E_C/h$ & $E_J/E_C$ & $\omega_{10}/2\pi$ & $\omega_{21}/2\pi$ & $\Gamma_{10}/2\pi$ & $\Gamma_{\phi}/2\pi$ & Ext. \\
\hline
1 & $12.7$ & $0.59$ & $21.6$ & $7.1$ & $6.38$ & $0.073$ & $0.018$ & $90\%$ \\
\hline
2 & $10.7$ & $0.35$ & $31$ & $5.13$ & $4.74$ & $0.041$ & $0.001$ & $99\%$ \\
\hline
3 & $-$ & $-$ & $-$ & $4.88$ & $4.12$ & $0.017$ & $0.0085$ & $75\%$ \\
\hline
\end{tabular}
\caption{\label{Params} Parameters for samples 1, 2 and 3.  All values in GHz (except for the extinction and $E_J/E_C$).}
\end{table}

The extinction efficiency of sample 2 is much better than sample 1. This is because sample 1 has a low $E_J/E_C\sim21.6$, which is barely in the transmon limit. For this value of $E_J/E_C$, charge noise still plays an important role as the energy band of the 0-1 transition is still dependent on charge \cite{Koch}. For sample 1 we find the charge dispersion is 7\,MHz and the dephasing is dominated by charge noise. By increasing $E_J/E_C$ to 31, we see much less dephasing in sample 2, which gives nearly perfect extinction of propagating resonant microwaves. Note that, the anharmonicity between $\omega_{10}$ and $\omega_{21}$ of sample 2 is close to $E_C$. This is not quite the case for sample 1 due to its low $E_J/E_C$ \cite{Koch}.

 \begin{figure*}
 \includegraphics[width=\columnwidth]{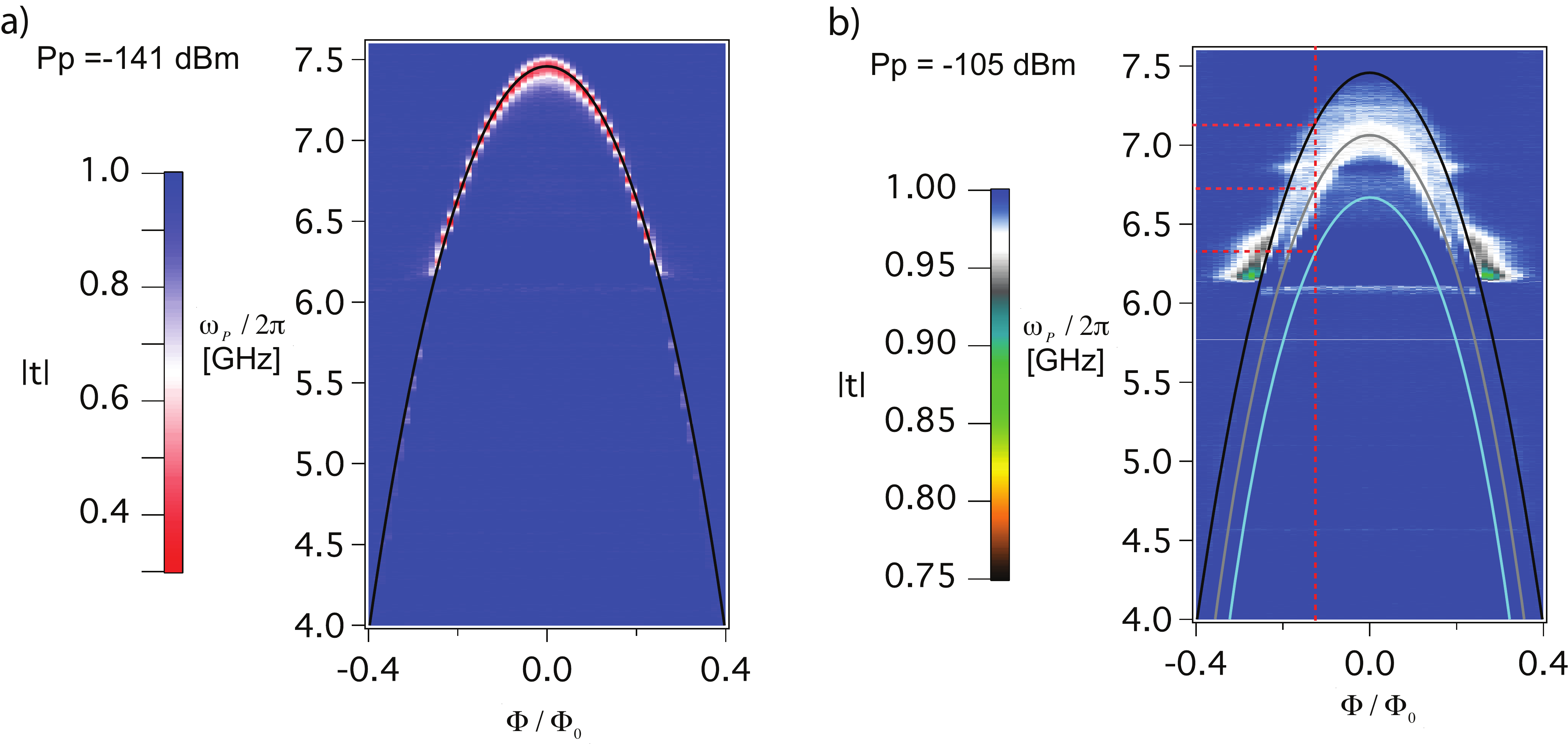}
\caption{ $|t|$ as a function of $\Phi$ for sample 1. a) At weak probe power, where $\Omega_p\ll\gamma_{10}$. The black curve is the theory fit to 0-1 transition. b) At high probe power, where $\Omega_p\gg\gamma_{10}$. The black and blue curve correspond to the 0-1 and 1-2 transition respectively.  The grey curve is the two-photon (0-2) transition. The red dash line indicates the flux bias point  and the corresponding $\omega_{10}$, $\omega_{20}/2$, $\omega_{21}$ for Fig. 2 and Fig. 4a, c and d. There is a stray resonance around 6.1 GHz.}
\end{figure*}

\section{Mollow Triplet and Aulter-Townes Splitting}
As we mentioned in the previous section, the transmon also has higher level transitions, in particular, we are interested in the 1-2 transition with frequency, $\omega_{21}$. By using 2-tone spectroscopy, the $\omega_{21}$ transition can be directly measured. We can saturate the $\omega_{10}$  transition by applying a pump field at $\omega_{10}=7.1$ GHz, and measure the transmission properties using a weak probe $\omega_{p}$. As the pump power is increased, the population of the first excited state increases, therefore, we start to observe photon scattering from the 1-2 transition, which appears as a dip in transmission at $\omega_{p}=\omega_{21}$, see Fig. 4a. The dip in transmission grows until the 0-1 transition becomes fully saturated. From this, we extract $\omega_{21}/2\pi=6.38$\,GHz for sample 1. Therefore, the two-photon (0-2) transition should be equal to 6.74 GHz, consistent with the observation in Fig. 3b. The linewidth of $\omega_{21}$ is around 120\,MHz, this dephasing mainly comes from the charge dispersion. Further increasing the pump power at $\omega_{10}$, we observe the well known Mollow triplet \cite{Astafiev1,Mollow} (Fig. 4b, sample 3). The Rabi splitting of the triplet can be used to calibrate the power at the atom. The Mollow triplet can be explained in the dressed state picture, where the two lowest levels split by the Rabi frequency. These four states give three different transitions, indicated by red, brown and blue arrows in the inset of Fig. 4b. Note that the way we observed the triplet here is different from that in \cite{Astafiev1}. We probe the transmission of these triplet transitions instead of looking at the emission spectrum. We see that the center transition is much less visible, because we pump at the frequency which saturates the transition.

 \begin{figure*}
 \includegraphics[width=\columnwidth]{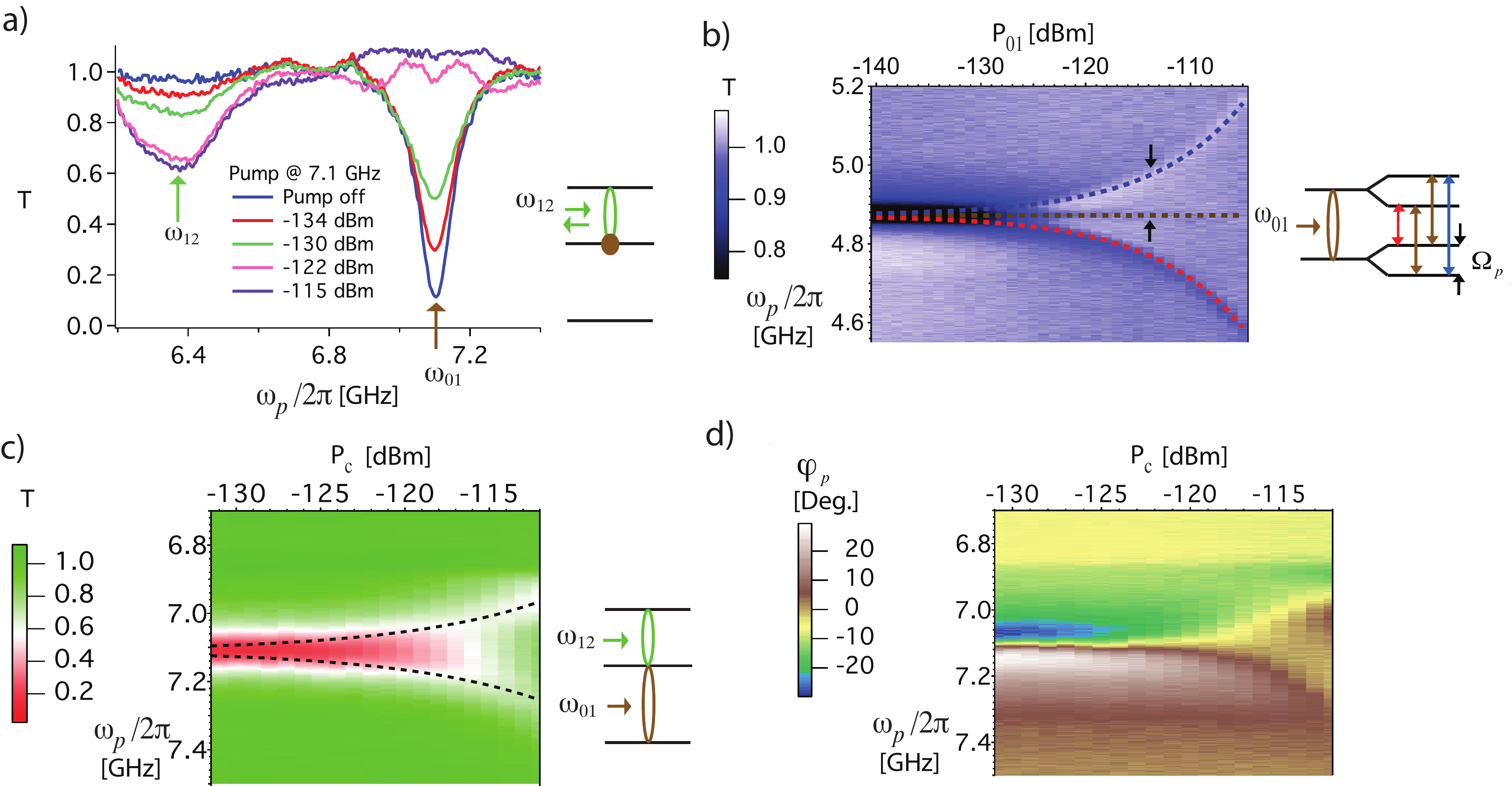}
\caption{ Two tone spectroscopy. a) As the frequency of a weak probe field is swept, a second microwave drive is continuously applied at $\omega_{10}$ with increasing powers. We see another dip gradually appears in the probe transmission response. b) $T$ as a function of probe frequency and pump power. As the power of  $\omega_{10}$ further increases, we see the Mollow triplet. The dashed lines indicate the calculated position of the triplet. (Inset) Schematic of triplet transitions in the dressed state picture. Note that we use sample 3, where $\omega_{10}/2\pi=4.88$ GHz. c) A second microwave drive is applied at $\omega_{21}$  with variable power, $P_c$. Magnitude response in (c). As $P_c$ increases, we see induced transmission at $\omega_p=\omega_{10}$. With a strong drive applied, the Aulter-Townes splitting appears with the magnitude of $\Omega_{c}/2\pi $ (Black dashed lines.) d) Phase response of the probe. }
\end{figure*}

With a weak resonant probe field, $\Omega_p\ll\gamma_{10},\omega_p=\omega_{10}$, and a strong resonant, $ \omega_c=\omega_{21}$, control field, the 0-1 resonance dip splits with the magnitude of $\Omega_c$ \cite{Abdumalikov}, this is known as the Aulter-Townes Splitting(ATS) \cite{Autler}. The magnitude and phase response are shown in  Fig. 4c and 4d respectively. In the magnitude response, we see that the transmon becomes transparent for the probe at $\omega_{p}=\omega_{10}$ at sufficiently high control power. In the phase response, we see the probe phase, $\varphi_p$, depends on the control power, $P_c$. 

In the following application section, we demonstrate two devices based on these effects which could be utilized in a microwave quantum network. By making use of the ATS,  we demonstrate a router for single photons.
By using the high nonlinearity of the atom, we demonstrate a photon-number filter, where we convert classical coherent microwaves to a nonclassical microwave field.

\section{Applications}
\subsection{The Single-Photon Router}
The operation principle of the single-photon router is explained as follows. In the time domain (see Fig. 5a), we input a constant weak probe in the single-photon regime, $\left \langle N_p \right \rangle\ll 1$, at $\omega_{p}=\omega_{10}$. We then apply a strong control pulse, around 30\,dB more than the probe power, at the $\omega_{21}$ frequency. When the control is off, the probe photons are reflected by the atom, and delivered to output port 1. When the control is on, the probe photons are transmitted based on ATS, and delivered to output port 2. We measure the reflected and transmitted probe power simultaneously in the time domain. This is crucial to investigate if the microwave photon transport is a fully coherent process, $i.e.$ the transmission dip seen in Fig. 2a is that the photons are being reflected (not due to dissipation). Note that we are phase sensitive since we measure $\left \langle V\right \rangle^2$ rather than $\left \langle V^2 \right \rangle$, this means that we are only sensitive to the coherent part of the signal. The experimental setup is shown in Fig. 5a.

 \begin{figure*}
 \includegraphics[width=\columnwidth]{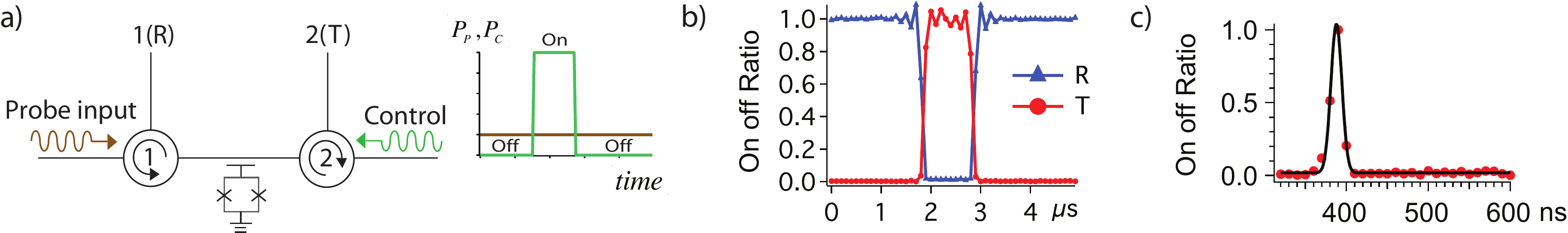}
\caption{The Single-Photon Router, data for sample 2. (a) Measurement setup and the control pulse sequence. A strong control pulse at  $\omega_{c}=\omega_{21}$ is used to route a weak continuous microwave $\omega_p=\omega_{10}$. Depending on whether the control pulse is on or off, the probe field is delivered to output port 2 or 1, respectively. (b), (c) Normalized on-off ratio (see text) of the transmittance (T) and reflectance (R) of $\omega_{p}$ measured simultaneously. b) the control pulse is shaped as a square pulse with $1\,\mu s$ duration. c) a gaussian pulse with a duration of $10\,\ ns$, we see up to 99$\%$ on-off ratio. The black curve in (c) is a gaussian fit to the data. }
\end{figure*}

The results are shown in Fig. 5b,c with two different control pulses for sample 2. In Fig. 5b (c), we use a square (gaussian) control pulse with a duration of 1 $\mu s$ (10 ns). As expected, when the control signal is on, the probe power of the transmitted signal is increased and we see a corresponding decrease in the reflected probe signal. $99\%$ probe on-off ratio is achieved in both reflection and transmission for sample 2. We also see that the on-off ratio does not depend on the control time. In Fig. 5c, the time resolution of our digitizer detector/arbitrary waveform generator is 5 ns, which prevents us from accurately measuring pulses less than about 10 ns. The response follows nicely with the waveform of the control pulse. The ringing appearing in Fig. 5b are artifacts of the digitizer. In the setup of Fig. 5a, we send $\omega_{10}$ and $\omega_{21}$ in opposite direction with respect to each other. We can also send pulses in the same direction by using a microwave combiner at one of the input ports and get the same results, as expected. Note that we use the on-off ratio here \cite{Hoi}, because it is not possible to do a full calibration of the background reflections in the line and leakage through circulator 1 [Fig. 5(a)].

The speed of our router sample 1(2) is predicted to be $1/\Gamma_{10}\sim$ 2 ns (4 ns). We show that the router works well down to the time limit of our instruments. By engineering the relaxation rate, it should be possible to achieve switching times in the sub nanosecond regime. In addition, the routing efficiency, $R=|r_0|^2$, can be improved by further reducing $\Gamma_{\phi}$. The improvement in sample 2 compared to sample 1 was achieved by increasing the $E_J/E_C$ ratio. This reduced the sensitivity to the charge noise and therefore the dephasing.

Our router can also easily be cascaded to distribute photons to many output channels. Fig. 6a shows 4 atoms (A,B,C,D) in series, each separated by a circulator. The $\omega_{10}$ of the atoms are the same, while the $\omega_{21}$ are different. This arrangement can be designed in a straightforward manner by controlling the ratio of $E_J/E_C$.  By turning on and off control tones at the various 1-2 transition frequencies of different atoms, we can determine the output channel of the probe field, according to the table of Fig. 6b. For instance, if we want to send the probe field to channel 4, we apply three control tones at $\omega_{21,A},\omega_{21,B}$,$\omega_{21,C}$. Note that regardless of the number of output channels, all the control tones and the probe tone can be input through the same input port. Theoretically, the maximum number of output channels depends on the ratio of the anharmonicity and the width of the 1-2 transition, $\gamma_{21}$. Thus, there is a trade off between efficiency and the number of outputs.

 \begin{figure*}
 \includegraphics[width=\columnwidth]{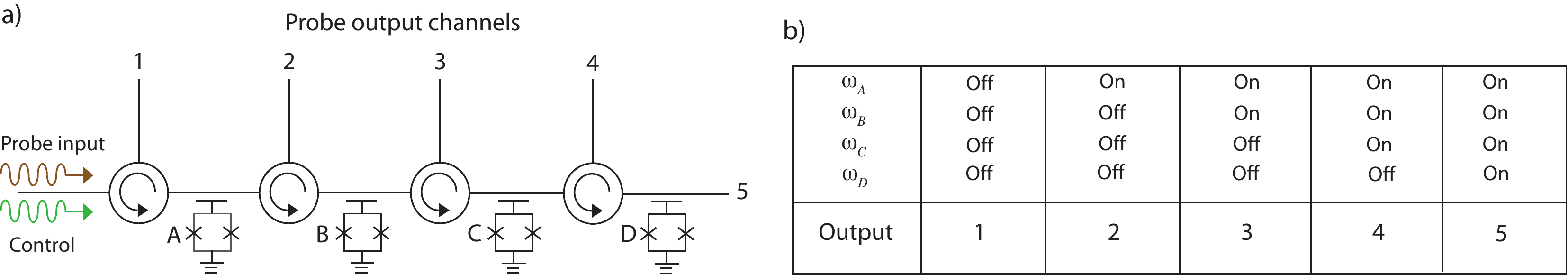}
\caption{Multiport router. (a) Cartoon of a multiport router: cascaded single-photon router to many output channels.  Here we show a 5 port router using 4 atoms (A,B,C,D) in series, each separated by a circulator. The $\omega_{10}$ of the atoms are the same, while the 1-2 transition frequencies, $\omega_{21,A}\neq\omega_{21,B}\neq\omega_{21,C}\neq\omega_{21,D}$, are different. By turning on and off control tones at the various 1-2 transition frequencies, we can determine the output channel of the probe field, according to the table b). }
\end{figure*}

\subsection{The photon-number filter}
In Fig.1b, we demonstrated the nonlinear nature of the artificial atom. This naturally comes from the fact that atoms can only reflect one photon at a time. To reveal the nonclassical character of the reflected field, we investigate the statistics of the reflected field. In particular, in this section, we show that the reflected field is antibunched \cite{Chang}. In addition, we also show that the transmitted fields is superbunched \cite{Chang}.

The incident coherent state can be written in terms of a superposition of photon
number states, with a poissonian distribution. For a weak probe field with $\left \langle N_p \right \rangle<0.5$, this coherent
field can be approximated using a basis of the first three-photon number states. For a
one-photon incident state, the atom reflects it, leading to
antibunching in the reflected field. Together with the
zero-photon state, the reflected field still maintains
first-order coherence, as there is a well defined phase between the zero and one photon states. Because the atom is not able to scatter more than one photon at a time, a two-photon incident state has a much higher probability of transmission, leading to superbunching in the transmitted field \cite{Zheng, Chang}. In this sense, our single artificial atom acts as a photon-number filter, which filters and reflects the one-photon number state from a coherent state. This process leads to a photon-number redistribution between the reflected and transmitted fields \cite{Zheng}.

A schematic illustration of the measurement setup is shown in Fig. 7a. This allows us to measure Hanbury Brown-Twiss \cite{HANBURY} type power-power correlations. We apply a resonant coherent microwave field at $\omega_{p}=\omega_{10}$, depending on whether we send the input through circulator 1 or 2, we measure the statistics of the reflected or transmitted field, respectively. The signal then propagates to a beam splitter, which in the microwave domain is realized by a hybrid coupler, where the outputs of the beam splitter are connected to two nominally identical HEMT amplifiers with system noise temperatures of approximately 7 K. We assume that the amplifier noise generated in the two independent detection chains is uncorrelated. After further amplification, the two voltage amplitudes of the outputs are captured by a pair of vector digitizers.

The second order correlation function \cite{Loudon} provides a statistical tool to characterize the field, it can be expressed as
\begin{eqnarray*}
g^{(2)}(\tau)=1+\frac{\left \langle \Delta P_1(t)\Delta P_2(t+\tau)\right \rangle}{[\left \langle P_1(t) \right \rangle - \left \langle P_{1,N}(t) \right \rangle][\left \langle P_2(t)\right \rangle-\left \langle P_{2,N}(t) \right \rangle]},
\end{eqnarray*}
where $\tau$ is the delay time between the two digitizers, $P_1,P_2$ are the output powers in ports 1 and 2, respectively, see Fig. 7a. $P_{1,N},P_{2,N}$ are the amplifier noise in ports 1 and 2 respectively, when the incident source is off. Therefore, $[\left \langle P_i(t)\right \rangle-\left \langle P_{i,N}(t) \right \rangle]$ represents the net power of the field from output port i, where $i=1,2$. $\left<\Delta P_1\Delta P_2\right>$ is the covariance of the output powers in ports 1 and 2, defined as $\left<(P_1- \left< P_1\right>)(P_2- \left< P_2\right>)\right>$.

 \begin{figure*}
 \includegraphics[width=\columnwidth]{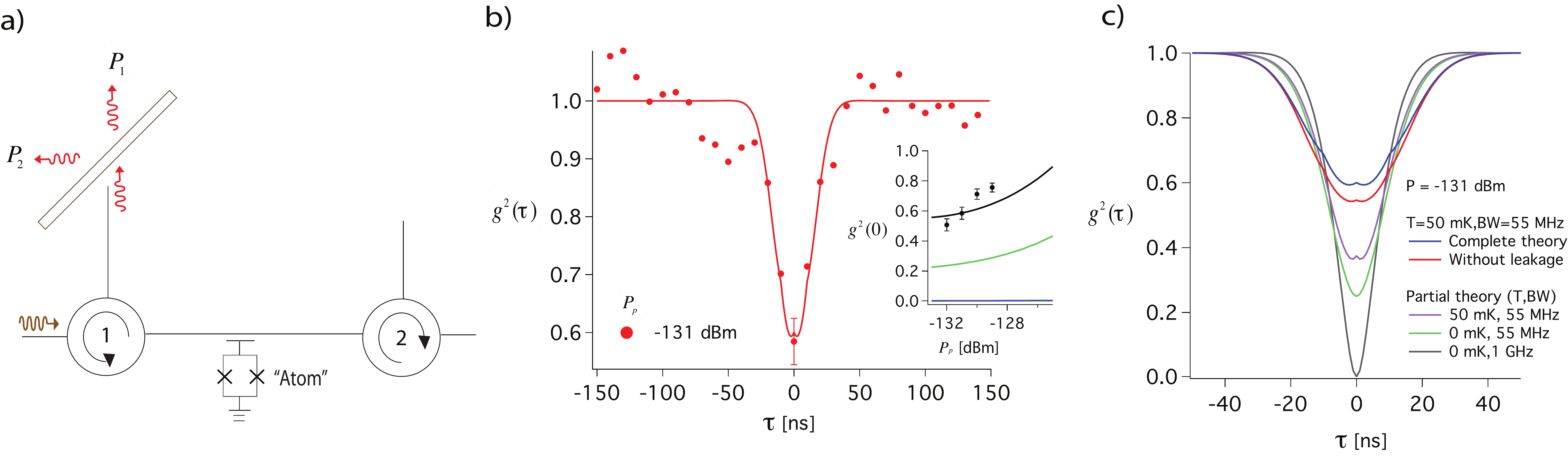}
\caption{Second-order correlation function of reflected fields generated by the artificial atom (sample 2). a) A schematic illustration of the physical setup, including circulators (labelled 1, 2) and the hybrid coupler which acts as a beam splitter for the Hanbury Brown-Twiss measurements \cite{HANBURY}. b) $g^{(2)}$ of a resonant reflected field as a function of delay time. We see the antibunched behavior of the reflected field. Inset: $g^{(2)}(0)$ as a
function of incident power. Complete theory including all four non-idealities: Black curve, BW= 55 MHz at 50 mK; partial theory including finite temperature and BW, but not leakage and trigger jitter (see text): Green curve, BW=55 MHz at 0 mK and Blue curve, BW=1 GHz at 0 mK. As the BW decreases or incident power increase, the degree of antibunching decrease. The error bar indicated for each data (markers) set is the same for all the points. c) Influence of BW, temperature, leakage and jitter on antibunching. The solid curves in b) and c) are the theory curves. For the curves with leakage, we assume the phase between the leakage field and the field reflected by the atom is $0.37\pi$. }
\end{figure*}

We had a trigger jitter of $\pm1$ sample between the two digitizers. To minimize the effect of this trigger jitter, we oversample and then digitally filter (average) the data in all the $g^{(2)}$ measurements. Here, the sampling frequency is set to $10^8$ samples/s with a digital filter with a bandwidth BW=55 MHz applied to each digitizer for all measurements. For a coherent state, we find $g^{(2)}(\tau)=1$ with the qubit detuned from $\omega_{10}$.

In Fig. 7b, we plot the measured $g^{(2)}(\tau)$ of the reflected field from our atom. At low powers, where $\left \langle N_p \right \rangle\ll 1$, we clearly observe antibunching of the field \cite{Chang}. The trace here was averaged over $2.4\times10^{11}$ measured quadrature field samples. The antibunching behavior at $P_p=-131$ dBm ($\left \langle N_p \right \rangle\sim0.4$), $g^{(2)}(0)=0.55\pm 0.04$, reveals the nonclassical character of the field. Ideally, we would find $g^{(2)}(0)=0$ as the atom can only reflect one photon at a time. The non-zero $g^{(2)}(0)$ we measured originates from four effects: 1) a thermal field at 50 mK temperature, 2) a finite filter BW, 3) trigger jitter between the two digitizers and 4) stray fields including background reflections in the line and leakage through circulator 1 [Fig. 7(a)]. The complete theory curves include all four non-idealities, the  partial theory curves include 1) and 2), but not 3), 4). The effects of these factors on our measured antibunching are shown in the theory plot Fig. 7c. For small BW, within the long sampling time, the atom is able to scatter multiple photons. If BW $\ll\Gamma_{10}, \Omega_p$, the antibunching dip we measure vanishes entirely. This interplay between BW and $\Omega_p$ yields a power dependent $g^{(2)}(0)$, as shown in the inset of Fig. 7b. In the ideal case, $i.e.$ for a sufficiently wide BW (1 GHz) at 0 mK, the theory gives $g^{(2)}(0)=0$, as expected.

In Fig. 8a, we see superbunching of the photons \cite{Chang} with $g^{(2)}(\tau=0)=2.31\pm 0.09>2$ at $P_p=-129$ dBm ($\left \langle N_p \right \rangle\simeq 0.8$) for the transmitted field. Superbunching occurs because the one-photon state of the incident field has been selectively reflected and thus filtered out from the transmitted signal, while the two-photon state is more likely transmitted. The three-photon state and higher number states are negligible. The transmitted state generated from our qubit is thus bunched even more than a thermal state, which has $g^{(2)}_{therm}(\tau=0)=2$. Fig. 8b shows the theoretical curves of $g^{(2)}(\tau)$ transmitted field under the influence of various effects. For the case of BW=1 GHz at 0 mK, indicated by the black curve, $g^{(2)}$ exhibits very strong bunching at $\tau=0$. At a later $\tau$($\sim$ 15 ns), $g^{(2)}$ for the transmitted field even appears antibunched \cite{Chang}, this is however not resolved in the experimental data. For the other curves, we see the degrading of superbunching due to the influence of BW, temperature and jitter. In Fig. 8c, we plot $g^{(2)}(0)$ as a function of incident
power, and clearly see the (super)bunching behavior decrease as the
incident power increases. For high powers, where $\left \langle N_p \right \rangle\gg1$, we find
$g^{(2)}(\tau)=1$. This is because most of the coherent signal then passes
through the transmission line without interacting with the qubit owing to saturation of the atomic response. We also plot the theoretical curves (Blue) at 0 mK for two different BW.

 \begin{figure*}
 \includegraphics[width=\columnwidth]{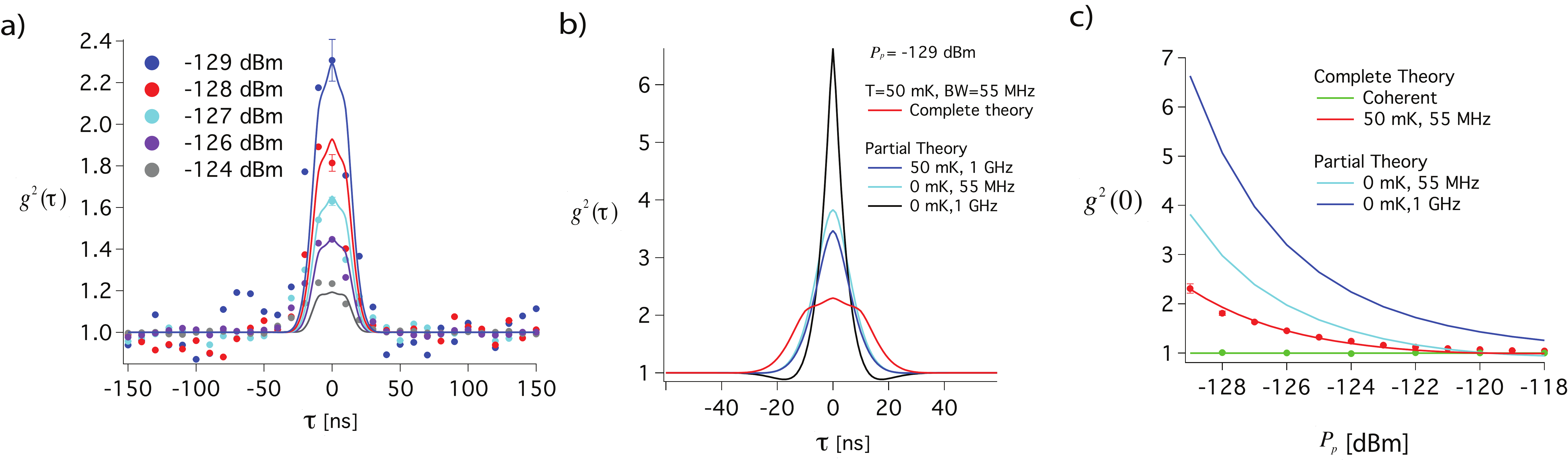}
\caption{Second-order correlation function of the transmitted fields generated by the artificial atom (sample 2). a) $g^{(2)}$ of the resonant transmitted microwaves as a function of delay time for five different incident powers. The peculiar feature of $g^{(2)}$ around zero in the theory curves in a) is due to the trigger jitter model. b) Influence of BW, temperature and jitter on superbunching. c) $g^{(2)}(0)$ of resonant transmitted field as a function of incident power. The result for a coherent state is also plotted. We see that the transmitted field
statistics(red curve) approaches that of a coherent field at high incident
power, as expected. For BW=1 GHz at 0 mK, we see very strong bunching at low incident power.
The error bar indicated for each data (markers) set is the same for all the points. The solid curves in a), b) and c) are the theory curves. For all measurements shown here we find, $g^{(2)}(\infty)=1$, as
expected. }
\end{figure*}

A single-mode resonator is used to model the digital filter. The theoretical curves in Fig. 7 and 8 are based on a master equation describing both the transmon and the resonator using the formalism of cascaded quantum systems \cite{Peropadre1}. The trigger jitter is modeled by the following: the value of $g^{(2)}(\tau)$ at each point is replaced by the average value of  $g^{(2)}(\tau$-10 ns), $g^{(2)}(\tau)$ and $g^{(2)}(\tau$+10 ns). We extract 50 mK from all these fits, with no additional free fitting parameters.

As we have shown, the single artificial atom acts as a photon-number filter, which selectively filters out the one-photon number state from a coherent state. This provides a novel way to generate single microwave photons \cite{Mallet,Bozyigit,Wilson4}.

\section{Discussion}
Microwave quantum optics with a single artificial atom opens up a novel way of building up a quantum network based on superconducting circuits. In such a system, superconducting processors can act as quantum nodes, which can be linked by quantum channels, to transfer flying photons (quantum information) from site to site on chip with high fidelity. In this way, the single-photon router can switch quantum information on nanosecond timescales and with 99\% efficiency, with the possibility of multiple outputs. The photon-number filter can act as the source of generation of flying microwave photons. These components have the advantage of wide frequency range compared to cavity-based systems \cite{Stojan,Bozyigit,Sandberg}. In addition, recent development of cross-Kerr phase shifter at the single-photon level based on superconducting circuit is also beneficial for a microwave quantum network \cite{Hoi3}.  

While microwave quantum optics with artificial atoms is a promising technology for quantum information processing, optical photons have clearly advantages on long distance quantum communication via a quantum channel. The development of hybrid quantum network would combine both advantages of these two systems. The early stages of optical-microwave interface have been demonstrated \cite{Kubo1,Kubo2,Matthias}, with other potential coupling mechanisms under investigation \cite{Kielpinski,Wang,Kim}.

\section{Summary}

Based on superconducting circuits, we study various fundamental quantum optical effects with a single artificial atom, such as photon scattering, Mollow triplet and Aulter-Townes Splitting. We further demonstrate two potential elements for an on chip quantum network: the single-photon router and the photon-number filter.

\section{Acknowledgments}
We acknowledge financial support from the EU through the ERC and the project PROMISCE, from the Swedish Research Council and the Wallenberg foundation. B.P. acknowledges support from CSIC JAE-PREDOC2009 Grant. We would also like to acknowledge Thomas M. Stace, Bixuan Fan, G. J. Milburn and O. Astafiev for fruitful discussions.

\section{References}


\end{document}